\documentclass[aps,nofootinbib,showpacs]{revtex4}
\usepackage[CJKbookmarks, pdftex, bookmarksnumbered, bookmarksopen, colorlinks, citecolor=blue, linkcolor=blue]{hyperref}
\usepackage[english]{babel}
\usepackage{amstext,amsmath,amssymb,amsfonts,bbm}
\usepackage[latin1]{inputenc}
\usepackage{graphicx}
\usepackage{epstopdf}
\usepackage{amsthm}
\usepackage{tocvsec2}
\usepackage{enumerate}
\usepackage{subfigure}
\usepackage{color}
\usepackage{tikz}
\usepackage[squaren,Gray]{SIunits}

\newcommand{\va}{\scriptscriptstyle}

\newcommand{\be}{\nopagebreak[3]\begin{equation}}
\newcommand{\ee}{\end{equation}}

\newcommand{\bee}{\nopagebreak[3]\begin{equation*}}
\newcommand{\eee}{\end{equation*}}

\newcommand{\ba}{\nopagebreak[3]\begin{eqnarray}}
\newcommand{\ea}{\end{eqnarray}}
\DeclareFontFamily{U}{rsfs}{}         
\DeclareFontShape{U}{rsfs}{m}{n}{<5> rsfs5 <6><7> rsfs7          %
  <8><9><10><10.95><12><14.4><17.28><20.74><24.88> rsfs10}{}     %
\DeclareMathAlphabet{\mathfs}{U}{rsfs}{m}{n}                     %
\newcommand{\mfs}[1]{\mathfs {#1}}                               %
\newcommand{\n}{{\nonumber}}

\newcommand{\sL}{{\mfs L}}

\newcommand{\sB}{{\mfs B}}

\newcommand{\sO}{{\mfs O}}

\newcommand{\x}{{ \bar x}}

\begin{document}

\title{ Energy-mass equivalence from Maxwell equations}

\author{Alejandro Perez}
\email{perez@cpt.univ-mrs.fr}
\affiliation{Aix Marseille Univ, Universit\'e de Toulon, CNRS, CPT, 13000 Marseille, France}
\author{Salvatore Ribisi}
\email{salvatore.ribisi@cpt.univ-mrs.fr}
\affiliation{Aix Marseille Univ, Universit\'e de Toulon, CNRS, CPT, 13000 Marseille, France}

\begin{abstract}
Since the appearance of Einstein's paper {\em``On the Electrodynamics of Moving Bodies''} and the birth of special relativity, it is understood that the theory was basically coded  within Maxwell's equations. The celebrated mass-energy equivalence relation, $E=mc^2$, is derived by Einstein using thought  experiments involving the kinematics of the emission of light (electromagnetic energy) and the relativity principle. Text book derivations  often follow paths similar to Einstein's, or the analysis of the kinematics of particle collisions interpreted from  the perspective of different inertial frames. All the same, in such derivations the direct dynamical link with hypothetical fundamental fields describing matter (e.g. Maxwell theory or other) is overshadowed by the use of powerful symmetry arguments, kinematics, and the relativity principle.  

Here we show that the formula can be derived directly form the dynamical equations of a massless matter model confined in a box (which can be thought of as a toy model of a composite particle). The only assumptions in the derivation are that the field equations hold and the energy-momentum tensor admits a universal interpretation in arbitrary coordinate systems. The mass-energy equivalence relation follows from the inertia or (taking the equivalence principle for granted) weight of confined field radiation. The present derivation offers an interesting pedagogical perspective on the formula providing a simple toy model on the origin of mass and a natural bridge to the foundations of general relativity. 
 \end{abstract}

\maketitle

\section{Introduction}

One of the striking results  of special relativity \cite{Einstein:2015:RSG} is  the implication of an equivalence between the concepts of inertia and energy. In one of his founding papers  \cite{original} Einstein arrives at the postulate of mass-energy equivalence by showing that a body emitting an energy $E$ via electromagnetic radiation will see its mass decreased by an amount $E/c^2$. Today textbooks give several different derivations. In the classic book \cite{Jackson:100964}, for instance, the equivalence is found by equating the force on a charged particle with its four-momentum variation. Another derivation is presented that uses the consistency with the relativity principle of the kinematics of colliding particles as seen from different inertial frames.  Perhaps the simplest (yet the most formal) derivation corresponds to 
the one that starts from the geometric (relativistic) free particle action 
\be
S[x(t)]=-m c \int dt \sqrt{|g_{\mu\nu} \dot x^\mu\dot x^\nu|}
\ee
whose non relativistic limit justifies the non-relativistic Lagrangian $L=m \dot x^2/2$, and in literally two lines of Hamiltonian analysis produces  the canonical Hamiltonian energy $E(v)=mc^2/(\sqrt{1-v^2/c^2})$ with $E(0)=mc^2$.

All these derivations are important and insightful in their own way and remain perhaps the simplest path to the equivalence formula.
However, none of these make the link between pure energy and mass dynamically explicit. They hide, in some sense, one  very important aspect which is perhaps the central one in view of  the necessary generalization of special relativity to include gravity in the general theory of relativity.  

The derivation proposed here is complementary to the standard account with the added value of presenting an explicit link between dynamics of a massless matter model---here the electromagnetic field, or a massless scalar field---and mass of its energy when confined in an idealized box. The derivation uses strongly the notion of stress-energy-momentum tensor of matter and the covariant interpretation of its physical content. The exercise paves the way for the understanding of the energy momentum tensor as the source of gravity in general relativity, and serves also as an introduction to the mathematics that is central in the definition of the theory.

We will consider radiation confined in an idealised box and we will show that this radiation confers inertia---encoded in a mass $m=E/c^2$---to the box, where $E$ is its energy content. The simplest model of radiation will be Maxwell electromagnetic fields which  strengthens the idea that special relativity is entirely encoded in the properties of electromagnetism (this was indeed the perspective adopted by Einstein in his 1905 revolution). As our derivation basically relies entirely on coordinate independence of the field equations and the conservation of the stress-energy-momentum tensor of the matter fields, the result should be valid for any generally covariant matter model. As an additional example we show that the construction also works for a massless scalar field. 

The idealized box confining the radiation can in turn be thought of as a poor's man model of a composite particle (such as a proton or a neutron): from modern computations \cite{Durr:2008zz} in quantum chromodynamics (QCD) we know that $99\% $ of the mass of a proton comes from the energy of the confined (not by a box but by the non-linear strong interaction) gluon-quark radiation while only the remaining $1\%$ is associated to the contribution of the rest mass of the quarks. The QCD confinement potential is replaced in our model by the box and the boundary conditions that require the fields to vanish at its walls \footnote{One could speculate that fundamental massive fermions like the electrons might be seen as confined massless radiation as well. Solutions of the Dirac equation can be interpreted as two massless Weyl fermions (the left handed and right handed components of the Dirac fermion) which due to the mutual interaction mediated by the mass term in the Dirac equation constantly annihilate into each other. In this process the momentum of each individual massless component bounces by changing the sign of the momentum in the direction of the spin (Schroedinger's {\em zitterbewegung} \cite{schrodinger1930kraftefreie}) as if confined in a box of a size of the order of the Compton wave length of the fermion. }.   

Our derivation could not realistically have replaced the historical one because it strongly uses the physical interpretation of the stress-energy-momentum tensor (a notion that even when present in the literature on Maxwell fields, became only central after the development of relativity) and the general covariance of the relativistic field equations (which also emerged with the understanding of general relativity). Nonetheless, all the mathematical ingredients and physical interpretation were arguably available in the context of Maxwell electromagnetism.  Yet the derivation we propose is straightforward only once modern tools and modern understanding of covariant methods are used (at the technical level, the derivation we present uses to a large extend the mathematical tools of general relativity: covariant derivatives and general coordinate invariance).  We hope that this paper will present the students with an alternative (perhaps technically more advanced) pedagogical perspective of both technical interest and conceptual value.   

The paper is organized as follows. In Section \ref{photonh} we give some motivation for our approach by using the heuristics of a photon trapped inside an accelerating box (or a box on a constant gravitational field from the perspective of the equivalence principle). In Section \ref{RIRI} we review the properties of Rindler coordinates which represent accelerating frames. In Section \ref{emc2-maxwell} we derive the formula $E=mc^2$  by analysing the energy content of an accelerating box trapping stationary electromagnetic radiation. In Section \ref{emc2-scalar} we do the same but for a massless scalar field which suggests the universality of our derivation.

In the Appendix we give supplementary material that answers some questions that naturally arise from the our analysis.  The only important piece of information for the proof of the main result of the paper, in Section \ref{emc2-maxwell}, is that the electric field in the rest frame of the box must be perpendicular to the boundary walls (which is obvious for an inertial box but requires justification for an accelerating one). In Appendix  \ref{maxwell-eq}
we briefly recall the structure of the covariant version of Maxwells equations and introduce its stress-energy-momentum tensor, we also calculate the properties of electromagnetic field at the boundary of an accelerating box of perfectly conducting walls. In Appendix \ref{work}, we show that the mass-energy equivalence formula can also be derived from the work done by an external agent accelerating the massless confined radiation. This is the analog of the heuristic argument using the photon given at the beginning of the paper.   We do the same for the scalar field in Appendix \ref{workscalar}.

\section{Heuristics with a trapped photon}\label{photonh}

\begin{figure}[hhhhhh]
\centerline{\hspace{0.5cm} \(
\begin{array}{c}
\includegraphics[width=7cm]{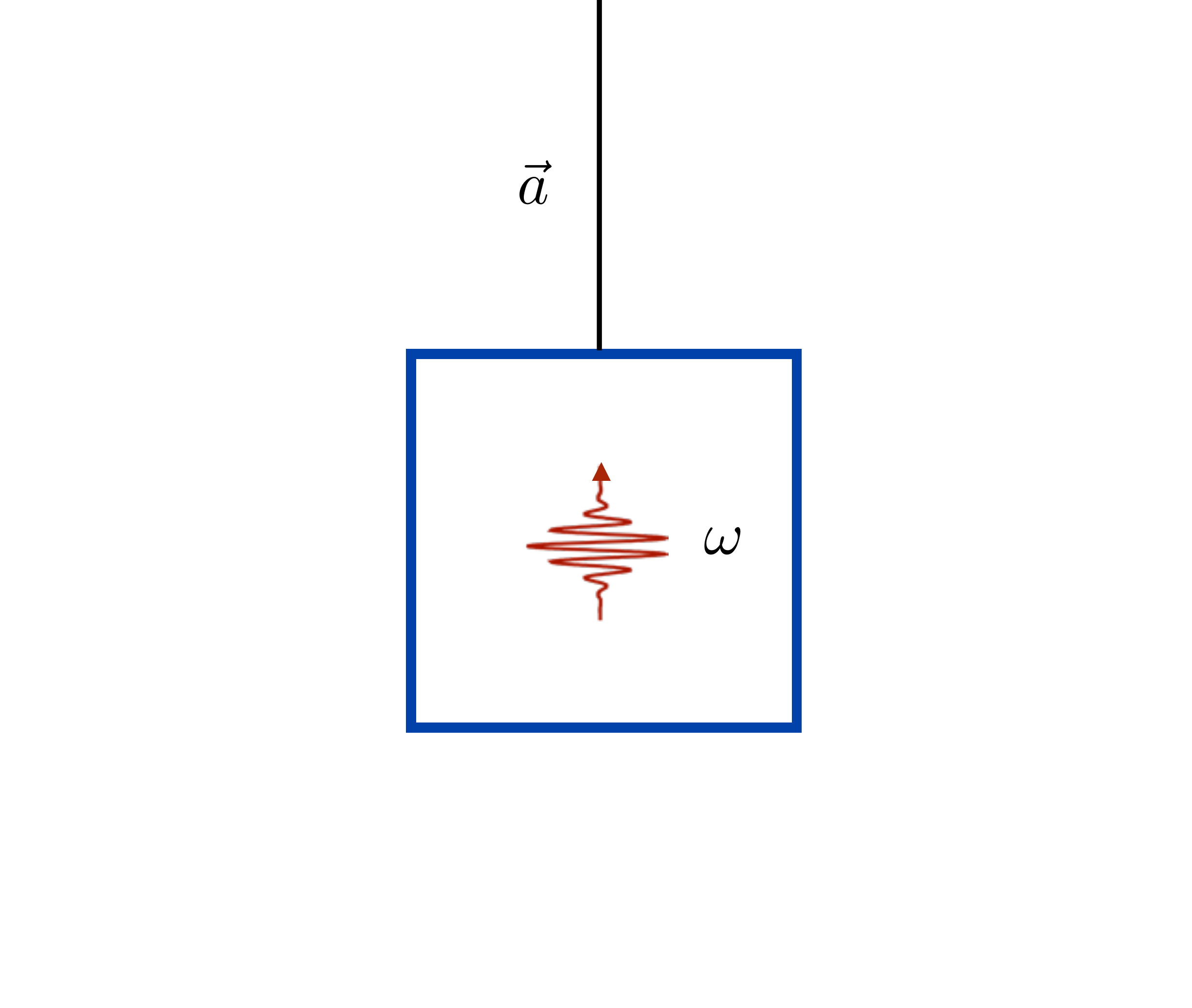} 
\end{array}\ \ \ \ \ \ \ \ \ \ \ \ \ \ \begin{array}{c}
\includegraphics[width=6.5cm]{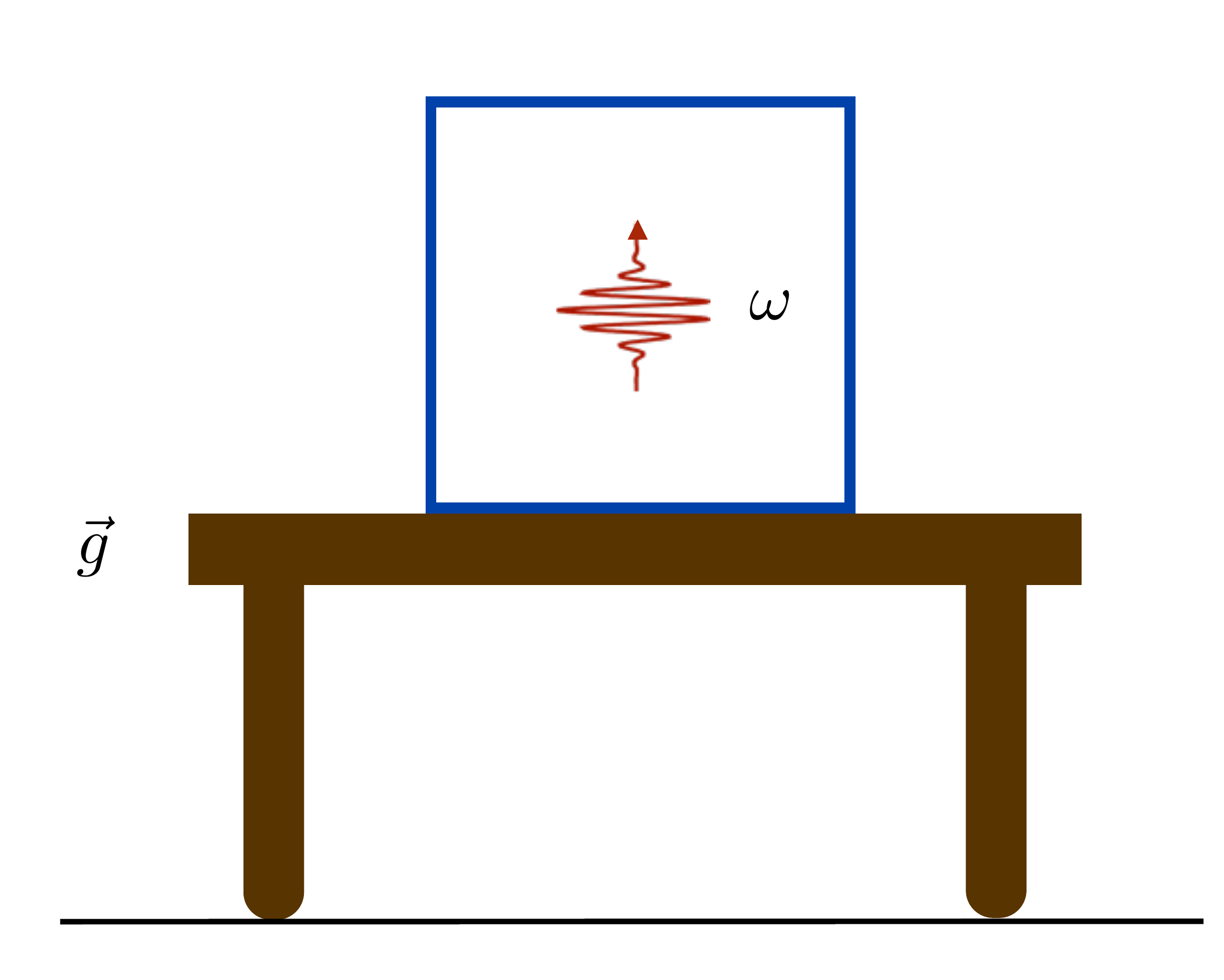} 
\end{array}\) } \caption{On the left panel: an accelerating box containing a photon. On average, a force $|F|=c^{-2} \hbar \omega |a|$ needs to be applied to maintain the acceleration. The inertia of the box is due to the average of the differential radiation pressure of the photon bouncing on the walls of the box (the pressure is smaller at the top than at the bottom due to the Doppler shift). On the right: the equivalent gravitational situation with the box on a table; the weight of the box is $W=c^{-2} \hbar \omega |g|$. Doppler shift is now replaced by the equivalent gravitational red-shift. }
\label{photon}
\end{figure}

As a warming up exercise let us first illustrate the basic idea by using the heuristics provided by the particle interpretation of electromagnetic radiation arising from quantum mechanics. Thus, consider a single photon of frequency $\omega$ trapped inside a cubic box of side $L$. Assume that the box is accelerated with acceleration $|a|$ in the upward direction (see Figure \ref{photon}). At time $t=0$ we also assume the box is at rest, and the photon is passing through the center of the box and moving up at the speed of light $c$. At time \ba t_1&=&\frac{L+|a| t_1^2}{2c}= \frac{L}{2c} \left(1+\sO\left(\frac{|a| L}{c^2}\right)\right)\n \\
&\approx& \frac{L}{2c}, \ea
the photon hits the top of the box.  We are assuming that the speed of the box when the photon hits the top is much smaller that the speed of light, hence $\frac{|a| L}{c^2}\ll 1$.
This can happen in two different limits: either the box is very small or the acceleration is very small with respect to the box size. 

When the photon hits the top its frequency in the rest frame of the top wall is $\omega_1=(1-|a|t_1/c) \omega$ due to the Doppler shift. Notice that this effect is expected from the non relativistic point of view also: in the limit where $\frac{|a| L}{c^2}\ll 1$ holds, the standard sound-wave-type Doppler red shift formula coincides with the (physically) correct relativistic one. Thus no explicit use of special relativity is being made here. Indeed, in the regime we are working, we can safely assume that time $t$ is always given by the very same inertial time (thought to be absolute time in pre-relativistic terms).  At time $t_1$ the momentum of the photon changes from $\hbar \omega_1/c\to -\hbar \omega_1/c$ in the vertical axis. We have that $\Delta p_1=-2\hbar \omega_1/c$.  

Now, the photon travels downwards and its frequency is back to $\omega$ (in the instantaneous rest frame of the box) thanks to the Doppler effect when it passes through the center. Indeed the frequency at the center is always $\omega$ as the photons energy must be stationary in the rest frame of the box. This is the easiest understood intuitively by considering the equivalent situation of the box on the table on the right panel of Figure \ref{photon}. This implies that at time $t_2$ the photon hits the bottom of the box with a local frequency $\omega_2=(1+|a| L/(2c^2)) \omega$ and $\Delta p_2=2\hbar \omega_2/c$. When the photon gets back to the center at time $t_3=2 L/c$ the average force $F=\Delta p/\Delta t$ is
\ba
|F|&=&\frac{\Delta p}{\Delta t}=\frac{2 \hbar \omega}{c} \frac{ \left(1+\frac{|a|L}{2c^2}-1+\frac{|a|L}{2c^2}\right)}{2 \frac{L}{c}} \\ \n 
&=&\frac{\hbar \omega}{c^2} |a|,
\ea
which implies that the box carries a mass $m=E/c^2$ with $E={\hbar \omega}$ (the quantum energy of the photon). The fluctuating character of the mass due to the bouncing back and forth of the photon would go away if we consider many photons in a suitable configuration that makes the radiation inside the box `stationary' in a way that will become precise in the following section. In the previous heuristic derivation, special relativity is of course hidden in the assumption that the momentum of a photon is $\hbar\omega/c$ but this is really only through quantum mechanics and the dynamical equations of Maxwell theory. That is indeed our point: the mass-energy equivalence relation is encoded in Maxwells dynamics.

The calculation can be improved by considering multiple photons in arbitrary configurations. The final answer remains the same although details become more and more cumbersome. The single photon example suffices as a motivation for what follows. The limitation of the present approach is the use of photons and quantum mechanics. Nevertheless, this simple argument also shows the heart of the reason for the energy-mass equivalence just derived; it resides in the structure of the photon momentum energy relation $p=\hbar \omega/c$ and $E=\hbar\omega$ which is also present in the classical theory, as realized by Einstein, and as pointed out in text books like the classic \cite{rindler2013essential}. More precisely, electromagnetic energy density $\rho\equiv(E^2+B^2)/(8\pi)$ and the electromagnetic Poynting vector---momentum density of the electromagnetic radiation---$P_i\equiv (\vec E\times\vec B)_i/(4\pi c)$ are such that for radiation (where $|E|=|B|$ and $\vec B\perp \vec E$) one has that $|\vec P|=E^2/(4\pi c)$ and $\rho=E^2/(4\pi)$ which confirms the energy-mass equivalence via the relationship $(energy/c^2)\times velocity=momentum$ for light. In what follows we will realize this in a clear-cut fashion by making appeal only to the structure of Maxwell equations without needing the details of a particular solution. As long as the radiation is confined inside the box in a stationary configuration  the energy-mass formula will follow.  

\section{Rindler coordinates}\label{RIRI}

We first need to get familiar with the description of an accelerated frame that will be used to represent those observers that are at rest with respect to the idealized box model of a composite particle made of confined electromagnetic radiation (or massless scalar field radiation).
 We will consider a box full of radiation (electromagnetic fields or massless scalar fields) in Minkowski spacetime whose metric in inertial coordinates takes the standard form
\be
ds^2=-c^2dt^2+dx^2+dy^2+dz^2.
\ee 
Inertial time translations define an isometry of flat spacetime so that $\xi^{\rm lab}\equiv \partial_{ct}$ satisfies a covariant equation known as the  Killing  equation \footnote{ The Killing equation follows from the fact that the Lie derivative of the metric along a vector field defining an isometry vanishes $\sL_{\xi^{\rm lab}} \eta_{ab}=2\nabla_{(a}\xi^{\rm lab}_{b)}=0$  \cite{Wald:1984rg}.}
\be\label{kilab}
\nabla_{(a}\xi^{\rm lab}_{b)}=0.
\ee
In order to describe the uniformly accelerating box in the $x$-direction  it will be convenient to introduce Rindler coordinates \cite{PhysRev.119.2082} that are related to the inertial coordinates  by 
\ba\label{inrin}
\n ct&=&\x \sinh(\tau)\\
\label{x} x&=&\x \cosh(\tau)
\ea
so that the flat metric becomes
\be
ds^2=-\x^2 d\tau^2+d\x^2+dy^2+dz^2.
\ee
These new coordinates are those associated with uniformly accelerated observers \cite{Wald:1984rg}. The inverse transformation is
\ba\label{riri}
\bar x&=&\sqrt{x^2-c^2t^2} \\
\tau&=&{\rm arctanh}\left(\frac{ct}{x}\right).
\ea
For later use it is important to write $\xi^a_{\rm lab}=\partial_{ct}^a$ in terms of Rindler coordinates, from \eqref{riri} we get
\ba\label{tritri}
\xi^a_{\rm lab}&=&\partial_{ct}^a=\frac 1c \frac{\partial\tau}{\partial  t} \partial^a_\tau+ \frac 1c\frac{\partial\bar x}{\partial  t} \partial^a_{\bar x}\n \\
&=&\gamma(\tau)\left(\frac{1}{\bar x} \partial^a_\tau-\frac{v(\tau)}{c} \partial^a_{\bar x}\right),
\ea
where we introduced the relativistic gamma factor $\gamma(\tau)=\cosh(\tau)=(1-\beta^2)^{-1/2}$ and $\beta(\tau)=\tanh(\tau)=v(\tau)/c$.
Also for later use, the 4-volume form (see for instance Appendix B in \cite{Wald:1984rg}) in terms of Rindler coordinates is 
\be\label{4v}
dv^{\va (4)}=\bar x \, d\bar x dy dz d\tau,
\ee
and the $3$-volume density for the simultaneity surfaces $\tau=$constant is
\be\label{v3}
d\Sigma_\tau=d\bar x dy dz.
\ee
Since the metric does not depend on $\tau$, 
\be\label{kiboxy} \xi^a_{\rm box}\equiv \partial_\tau^a=\bar x\cosh\tau\partial_{ct}^a+\bar x\sinh\tau\partial_x^a\ee is a Killing vector too
\be\label{kibox} \nabla_{(a}\xi^{\rm box}_{b)}=0.
 \ee 
 The subindex "box" is natural due to the fact that this Killing field is associated with the time translation invariance of the uniformly accelerating observers at rest with 
 the box that will contain the confined radiation as described in the following section (the isometry corresponding to this Killing field is the one associated with the invariance of the flat Minkowski metric under boosts).
 
 The four velocity $u^a_{\rm box}$ of these observers is just proportional the Killing vector, i.e.,  given by $u^a_{\rm box}=\xi_{\rm box}^a/{|\xi_{\rm box}|}$, explicitly
 \ba\label{ubox}
 u^a_{\rm box}&=&\cosh\tau\partial_{ct}^a+\sinh\tau\partial_x^a\n \\
 &=&\gamma(\tau)\partial_{ct}^a+\gamma(\tau)\beta(\tau) \partial_x^a.
 \ea 
The previous expression of $u^a_{\rm box}$ tells us that the stationary observers following the killing trajectories of \eqref{kiboxy} correspond to orbits of boosts in the $x$-direction with rapidity given by $\tau$.  One can easily compute their acceleration and find that it is constant (independent of $\tau$) and  given by
\ba\label{acele}
a_a^{\rm box}&=&c^2 u^b_{\rm box}\nabla_bu_a^{{\rm box}}=c^2 \frac{\xi^b_{\rm box}}{|\xi_{\rm box}|}\nabla_b\left(\frac{\xi_a^{\rm box}}{|\xi_{\rm box}|}\right) \n \\
&=&c^2 \nabla_a \log(|\xi_{\rm box}|)= c^2 \frac{d\x_a}{\x},
\ea
where to get the final line we have used the Killing equation \eqref{kibox}.
Therefore, \eqref{ubox}  defines the four velocity field of a box where its bulk points move along constant acceleration trajectories with 
\be\label{ace} |a_{\rm box}|=\frac{c^2}{\bar x}\ee 
Notice that even when all points of the box move at the same speed (the box behaves as a rigid box) different points have different acceleration, e.g. the bottom of the box and the top of the box accelerate differently so that the box remains un-stretched (the distance between the top and the bottom of the box remains fixed). This might be surprising at first sight but it is one of these counter intuitive facts in Lorentzian geometry. All the same, when the concept of acceleration of the box will be needed (only in the material presented in the Appendices) we will work under the assumption that $\frac{|a_{\rm box}| L}{c^2}\ll 1$ (already used in Section \ref{photonh}) in which case a single constant notion of acceleration can be assigned in an approximate manner to the whole box.  
 \begin{figure}
\centerline{\hspace{0.5cm} \(
\begin{array}{c}
\includegraphics[width=12cm]{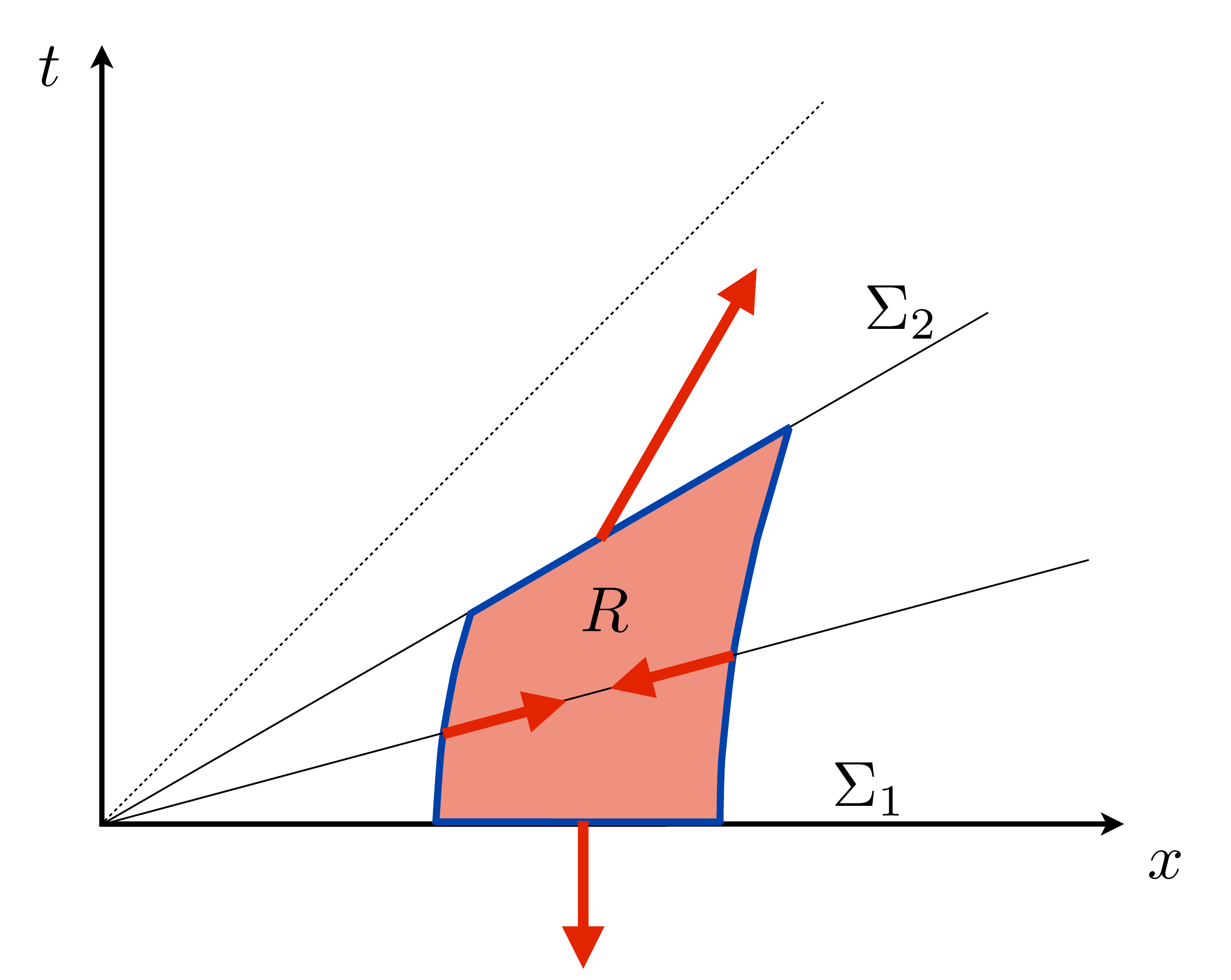} 
\end{array}\) } \caption{Accelerated box in Minkowski space-time. $\Sigma$ represent constant time surfaces while $R$ is the region inside the box. The hyperbola correspond to the trajectory of the walls perpendicular to the motion. }
\label{fig1}
\end{figure}

\begin{figure}
\centerline{\hspace{0.5cm} \(
\begin{array}{c}
\includegraphics[width=7cm]{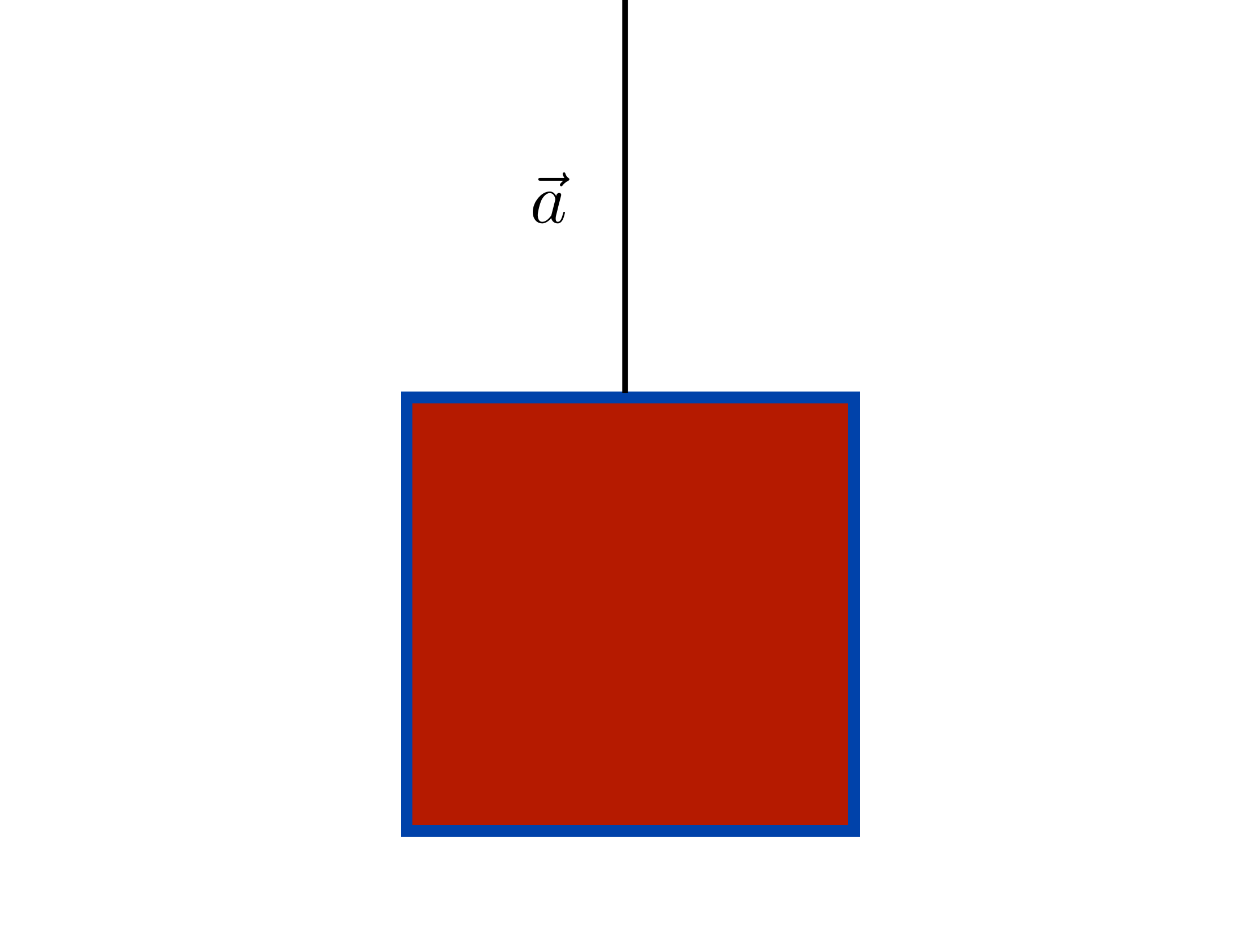} 
\end{array}\ \ \ \ \ \ \ \ \ \ \ \ \ \ \begin{array}{c}
\includegraphics[width=6.5cm]{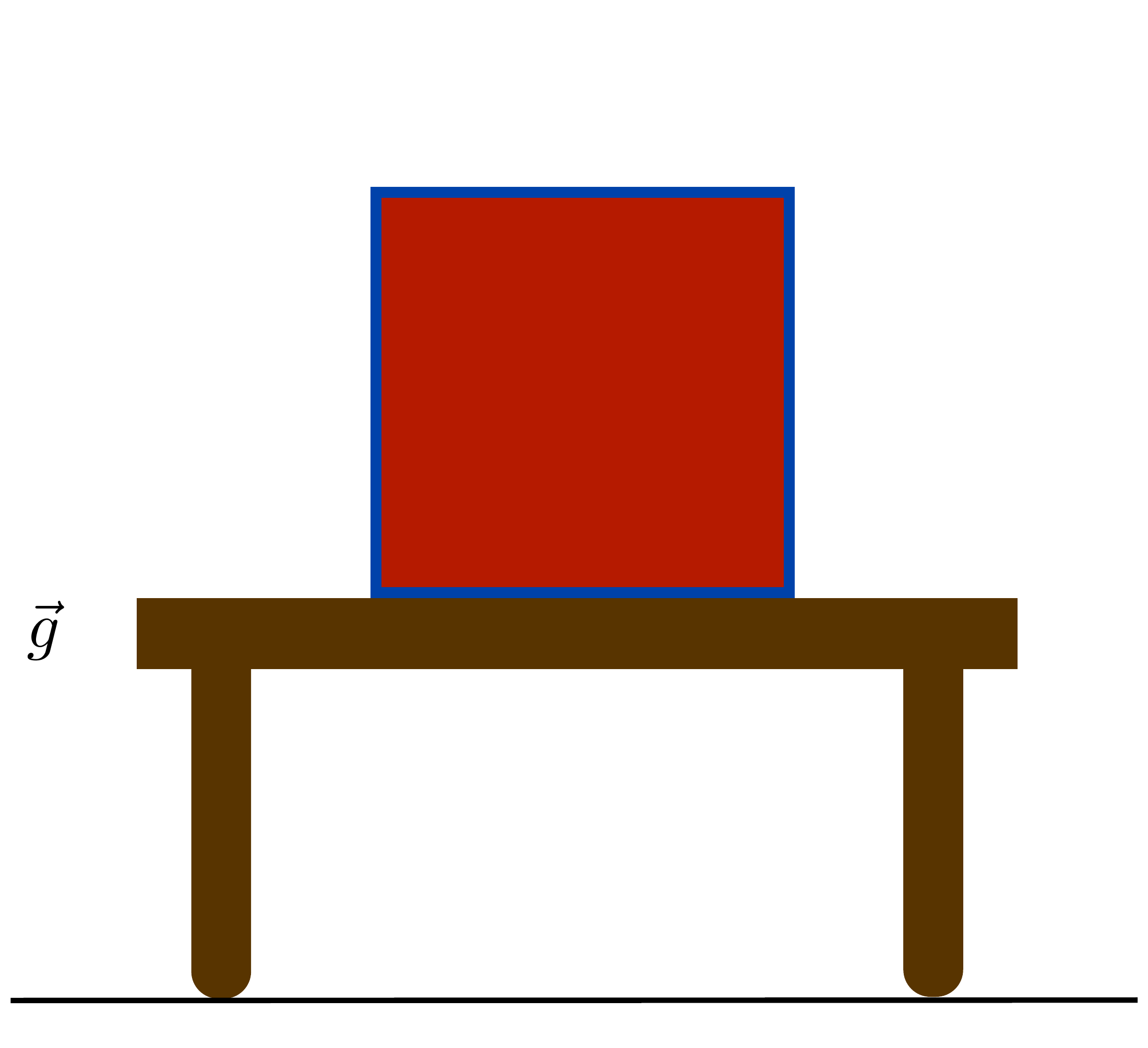} 
\end{array}\) } \caption{A box of accelerated radiation on the left. On the right the equivalent situation of radiation in a box near the vicinity of the earth. The energy-mass equivalence holds for the confined radiation. The only assumption is the stationarity of the radiation in the rest frame of the box and the validity of the field equations. }
\label{photon2}
\end{figure}

\section{The mass of Maxwell fields confined in a box}\label{emc2-maxwell}

In this section we derive the mass-energy equivalence formula from the inertial properties of a box made of perfectly conducting walls 
filled with electromagnetic radiation. We consider Maxwell theory and its solutions in Rindler coordinates, introduced in the previous section, and impose the boundary conditions that represent the presence of perfectly conducting walls in that frame. However, we will see that very little  information about the solutions is needed (and this is one of the nice features of the result). More precisely, the only explicit thing that enters the proof below  is that the electric field, in the rest frame of the accelerating box, must be perpendicular to the walls of the box. This is well known for a box in inertial motion, the fact that it remains true in the uniformly accelerated case is perhaps physically clear but technically less obvious. The proof is given in the Appendix \ref{piripipi} whose main results are discussed in  \ref{conse}, and the basic mathematical reason is that Maxwell's equations maintain very much the same structure on the accelerated frame as in an inertial one. 

Let us now compute the energy content of the box, as measured in the laboratory frame, at a given simultaneity surface of constant Rindler time $\tau$ (see Figure \ref{fig1}).
In order to do this we introduce the stress-energy-momentum tensor (energy-momentum tensor from now on) $T_{ab}$ of the Maxwell field and the current
\be
j^{\rm lab}_a\equiv -T_{ab}\xi_{\rm lab}^b.
\ee
The energy momentum tensor for electromagnetism is given in \eqref{jijo}; however, its explicit form is not important at the moment. 
By definition of the current in the previous equation, the energy content of the box of confined radiation at a given $\tau$ as measured in the lab frame is \be\label{primera}
E_{\rm box}(\tau)=-{c^2} \int_{\Sigma_\tau} j^{\rm lab}_a n^a \ d\Sigma_\tau,
\ee
where $d\Sigma_\tau= d\bar x dy dz$ is the volume density of the hypersurface $\tau=$constant as introduced in \eqref{v3} and $n^a$ is the normal to these hypersurfaces.
Explicitly,  $n^a=\bar x^{-1}\partial^a_\tau$ which when replaced in the previous equation gives\ba\label{energy1}
E_{\rm box}(\tau)&=&-{c^2}\int_{\Sigma_\tau} j^{\rm lab}_a n^a d\Sigma_\tau={c} \int\bar x^{-1} T_{t \tau}  d\bar x dy dz .\ea
Now, from \eqref{tritri} we get 
\ba\label{energy12}
E_{\rm box}(\tau)
&=&\gamma(v) c^2 \int_{\Sigma_\tau}\bar x^{-2}T_{\tau\tau}d\bar x dy dz -v {c}\gamma(v) \int_{\Sigma_\tau}\bar x^{-1}T_{\tau\bar x}d\bar x dydz.
\ea
As we show below, the second term in the previous equation vanishes if we demand that the radiation inside the box is stationary in its rest frame,  namely that the space components of the total linear momentum of the radiation vanishes in the rest frame of the box. Strictly speaking we only need the total momentum in the $\x$ direction to vanish. However, assuming we want our box to represent a simple model of composite particle that could be accelerated in arbitrary directions then the vanishing of all component of the space part of the linear momentum in the rest frame of the box will be a natural demand. In order to show that this is equivalent to the vanishing of the second term in \eqref{energy12}, let us now consider the energy current associated to the frame in which the box is at rest. So we have
\be\label{cubox}
j^{\rm box}_a=-T_{ab}u_{\rm box}^b=-\bar x^{-1}T_{ab}\partial_\tau^b,
\ee
where $u^a_{\rm box}=\xi^{a}_{\rm box}/|\xi_{\rm box}|=\bar x^{-1}\partial_\tau$ as given in \eqref{ubox} is the four velocity of the box. The momentum density of the radiation inside the box along the direction $\partial_{\bar x}$  as measured by the observer $u^a_{\rm box}$  at a time $\tau$ is given by $p_{\hat x}(\tau)=j^{\rm box}_a\partial_{\bar x}^a$. 
The condition for the vanishing of the total momentum of the radiation in the rest frame of the box in the $\x$ direction takes the form
\footnote{Notice that as the previous integral involves the notion of `$\x$-direction' at different point in the box, one would 	need in principle to parallel transport all these vectors at some reference point to be able to integrate (sum up) all the contributions. It is easy to check that such parallel transport is trivial in this case.}
\be\label{REST}
\int_{\Sigma_\tau} j^{\rm box}_a \partial_{\bar x}^a dv^{\va (3)}=-\int_{\Sigma_\tau}\bar x^{-1}T_{\tau\bar x}d\bar x dydz=0.
\ee
The previous is the only real requirement on the solutions of the electromagnetic field inside of the confining box. It has a natural physical meaning corresponding to restricting  the radiation to  a stationary configuration which reflects the notion of a compact particle-like object that we have in mind. In more general situations the two terms in \eqref{energy12} are important.

The previous stationarity condition reduces  \eqref{energy12}  to
\ba\label{ultima}
E_{\rm box}(\tau)
\label{etau}&=&\gamma(\tau)\int_{\Sigma_\tau}\frac{c^2}{\bar x^{2}}T_{\tau\tau}d\bar x dydz
\ea 
Now we show that the integral in the previous equation does not depend on $\Sigma_\tau$. A simple calculation of the divergence of the current \eqref{cubox} yields
\ba\label{ultima+}
\nabla^a j^{\rm box}_a=-\nabla^a\left(\bar x^{-1}T_{ab}\xi_{\rm box}^{b}\right)
&=&
\bar x^{-1}\sigma^a E^{\rm box}_{a} \delta_{\rm box} -T_{a\tau}g^{ac}\nabla_c\bar x^{-1}\n\\
&=&\bar x^{-2}T_{\tau\bar x}
\ea
where in the first line we have used \eqref{kibox} and (\ref{divj}), and in the second line the fact that $\sigma^a E^{\rm box}_{a}=0$ (the electric field in the box frame must be orthogonal to the surface current for perfectly conducting walls).  Indeed, the electric field $E^{\rm box}_{a}$ at the walls of the box is orthogonal to the walls of the box while the normal component of the magnetic field $B^{\rm box}_{a}$ vanishes at the walls.  Even though this might be physically clear, the mathematical proof from Maxwell equations is tricky because one is on a non-inertial frame.  We present it in the Appendix \ref{piripipi} and \ref{conse}. 

As implied by \eqref{ultima+} the current $j^{\rm box}_a$ is not locally conserved; nevertheless, when we integrate it in space-time region $R$ swept by the box (see figure \ref{fig1}) we find
\ba\label{primerita}
\int_{R} \nabla^aj_a^{\rm box}dv^{(4)}
&=&\int d\tau \left(\int_{\Sigma_\tau}\bar x^{-1}T_{\tau\bar x}d\bar x dydz\right)=0,
\ea
where we used \eqref{4v} and the last integration vanishes because the quantity in the parenthesis vanishes due to the stationarity condition \eqref{REST}.
Now, it follows from the perfect conductor boundary conditions that
\be
\left. j^a_{\rm box}N_a\right|_{\rm walls}=0,
\ee
where $N^a$ is the normal to the walls of the box (see detail proof in Appendix \ref{conse}, equation \eqref{ortoto}).
As a result, Gauss theorem implies that the flux across the boundary of the region $R$ receives only contributions from the spacelike components of the boundary of $R$ (see in Figure \ref{fig1}), namely
\be\label{quelop}
0=\int_{R}\nabla^aj_a^{\rm box}dv^{(4)}=\int_{\Sigma_2}j_a^{\rm box}u_{\rm box}^a d\Sigma_2-\int_{\Sigma_1}j_a^{\rm box}u_{\rm box}^ad\Sigma_1,
\ee
hence the integration of $j_a^{\rm box}u_{\rm box}^a$ does not depend on the $\tau=$constant hypersurface $\Sigma_\tau$: it is a constant of motion.
What is the physical interpretation of that constant? From the fact that $u^a_{\rm lab}=u^a_{\rm box}$ at $\tau=0$ we see that this constant is nothing else but the rest energy of the radiation in the box
\ba\label{energia}
E(0)=-\int_{\Sigma}c^2 j_a^{\rm box}u_{\rm box}^a d\Sigma=\int_\Sigma c^2 T_{ab}u_{\rm box}^au_{\rm box}^bd\Sigma=\int_\Sigma c^2 \bar x^{-2}T_{\tau\tau}d\bar x dydz.
\ea
Therefore, equation \eqref{ultima} takes the form
\be
{E_{\rm box}({v})=\gamma(v)E_{\rm box}(0)},
\ee
where we are now using the direct correspondence $\tau={\rm arctanh}(v/c)$ and hence trading $\tau$ by the velocity $v$ in the previous expression. Expanding $\gamma(v)$ to leading order on $v$ we find
\ba
E_{\rm box}(v)=\left(1+\frac{v^2}{2c^2}\right)E_{\rm box}(0)+\sO\left(\frac{v^4}{c^4}\right) E_{\rm box}(0).
\ea
Correspondence with the non relativistic limit requires 
\be \label{res}\boxed{E_{\rm box}(0)=mc^2}\ee
where $m$ is the rest mass of the confined electromagnetic radiation.

\section{The massless scalar field case}\label{emc2-scalar}

The result of the previous section depends of the field equations only and in a generic manner, in the sense that no particular solutions need to be considered for the proof. The first important ingredient is the conservation of the energy-momentum tensor in the bulk of the box (which follows from the validity of the field equations). The second is the behaviour of the divergence of $j_a^{\rm box}$ in \eqref{ultima+} where the reflecting boundary conditions constrain the electric field (as measured in the box frame) to be orthogonal to the walls, and the third is the orthogonality of $j_a^{\rm box}$ to the walls. Both these ingredients follow from the structure of the field equations at the boundary as shown in Appendix \ref{conse}. But the important things is that no specific solution needs to be chosen to prove these properties: these are generic consequences of the equations and the physical conditions at the idealized walls of the box. Therefore, one would expect that the proof of the previous section should be valid (with small adjustments) for any massless matter model. We do not have a general proof of this; nevertheless, we can at least build up evidence by exhibiting another simple example: the massless scalar field.

The field equation of a  massless scalar field $\phi$ is 
\be\label{fe}
\square \phi\equiv g^{ab}\nabla_a\nabla_b \phi= j,
\ee
where $j$ is a source term (necessary to impose the boundary conditions at the walls of the box).
The stress-energy-momentum tensor is given by 
\be\label{emt}
T_{ab}=\nabla_a\phi \nabla_b\phi-\frac{1}{2} g_{ab}\ g^{cd} \nabla_c \phi \nabla_d \phi.
\ee
Direct calculation of the divergence of the stress-energy-momentum tensor (\ref{emt})  yields
\be\label{divtabp}
\nabla^aT_{ab}=j \nabla_b\phi,
\ee
which in the absence of sources vanishes identically.
When the radiation is confined inside a box, made of perfectly reflecting walls (as in the Maxwell case) surface charges appear. We write
\be j=\sigma \delta_{\rm box},\ee
where $\sigma$ represent the surface charge density.
This is the analog of the surface electric charges and surface current in a perfect conducting wall in electromagnetism. They are the responsible of imposing reflecting boundary conditions that, in the present case, boil down to $\phi=0$ at the box walls. 
These sources fix the normal derivative of the scalar field: from \eqref{fe}, and the Gauss law applied to the vector field $\nabla^a\phi$, it follows that \be\label{nono} N^a\nabla_a \phi=\sigma. \ee
or simply 
\be \label{chichita}\nabla_a \phi=\sigma N_a. \ee
As in the case of Maxwell fields we start form the definition of the energy content of the box in the lab frame \eqref{primera}.
The argument follows the same lines from \eqref{primera} to \eqref{ultima} where the field equations do not really enter. Things change slightly when considering the current 
\eqref{cubox} whose divergence remains 
\ba\label{ultima+s}
\nabla^a j^{\rm box}_a=-\nabla^a\left(\bar x^{-1}T_{ab}\xi_{\rm box}^{b}\right)
&=&
-\bar x^{-1} \sigma (u^a_{\rm box}\nabla_a\phi) \delta_{\rm box} -T_{a\tau}g^{ac}\nabla_c\bar x^{-1}\n\\
&=&\bar x^{-2}T_{\tau\bar x},
\ea
due to the fact that $u^a_{\rm box}\nabla_a\phi=\bar x^{-1} \partial_\tau \phi=0$ at the boundary (either because $\phi=0$ for all $\tau$ or, equivalently, due to equation \eqref{chichita} and the fact that $u_{\rm box}^a  N_a=0$). Therefore, the equivalent of equation \eqref{primerita} is also valid for the scalar field as long as the stationarity condition \eqref{REST} is satisfied for the scalar field inside the box. Now, the validity of equation \eqref{quelop} depends on the validity of $j^{\rm box}_aN^a=0$. Using the definition of the energy-momentum tensor and the fact that  $u_{\rm box}^a  N_a=0$ we see that
\be
\left.j^{\rm box}_a N^a\right|_{\rm walls}=-(u_{\rm box}^a\nabla_a\phi)(N^b\nabla_b\phi)=0,
\ee
due to the boundary condition  $u^a_{\rm box}\nabla_a\phi=\bar x^{-1} \partial_\tau \phi=0$. The rest of the argument from \eqref{quelop} to the main result \eqref{res} 
now follows exactly as in the Maxwell case. 

\section{Conclusions}

We have shown that confined radiation in an idealized box with walls imposing perfectly reflecting boundary conditions for both Maxwell electromagnetic fields and massless scalar fields has an inertial mass given by its energy content divided by the square of the speed of light. Our calculation relies entirely on general properties of the solution of the field equations and the properties of the energy momentum tensor of the confined radiation. The only explicit requirement on the solutions is that the radiation be in a stationary state of vanishing total linear momentum in the frame of the box (not moving inside the box). This assumption is compatible with the idea of the box representing a toy-model of a composite particle (an ultra simplified classical model of  proton or a neutron). The calculations done in this solvable simple model of a composite particle has deep conceptual implications making natural the possibility that all mass parameters in our physical models could have a more fundamental description in terms of more basic degrees of freedom (mass as an emergent notion). 

The proof of main claim is straightforward once the relevant equations are written in covariant form. Even when no gravitational field is invoked, the result follows naturally from the application of the mathematics of general relativity. On the physical front, the naturalness of the energy-momentum tensor as the source of gravity is made more transparent by our pre-gravitational analysis. For those reason we expect the paper to be useful from a pedagogical perspective.

\section{Acknowledgements}

The basic idea of this paper started in discussion with a group students of the first year master in theoretical physics at Aix-Marseille University. 
We thank discussions with F. Balfour, M. L. Frisch Sbarra, A. Vesperini, and S. Charfi.

\begin{appendix}

\section{Maxwell equations and boundary conditions}\label{maxwell-eq}
Maxwell equations in covariant form and in the presence of sources are
\ba\label{me}
\nabla^a F_{ab}&=&-{4\pi} J_b\label{m1}\\
\nabla_{a} F_{bc}+\nabla_{b} F_{ca}+\nabla_{c} F_{ab}&=&0\label{m2}.
\ea
where $F_{ab}=-F_{ba}$ is the electromagnetic field strength, and $J_a$ is the electric four-current. For an arbitrary observer with four velocity $u^a$ the electric field is given by \be E_a=F_{ab}u^b\ee while the magnetic field is \be B_a=-\frac12\epsilon_{abcd} F^{cd} u^b. \ee 
The stress-energy-momentum tensor of the electromagnetic field 
is 
\be\label{jijo}
T_{ab}=\frac{1}{4\pi}(F_{ac} F^{\ c}_{b}-\frac{1}{4} g_{ab} \ F_{cd} F^{cd}).
\ee
It follows from the validity of \eqref{m1} and \eqref{m2} that the divergence of  $T_{ab}$ is given by 
\be\label{a6}
\nabla^aT_{ab}=J^a F_{ab}.
\ee
We will assume that the electromagnetic field is confined in a box without charges inside. Therefore, $J^a=0$ in the bulk of the box.
However, boundary currents must be present to ensure that the fields vanish outside the confining box (they are responsible for enforcing reflecting boundary conditions).  
We assume that the walls are made of a perfect conductor with infinitely light charge carriers that can move freely.
We will hence write the current as 
\be
J^a=\sigma^a \delta_{\rm box}, 
\ee
where $\sigma^a$ is the surface current and $\delta_{\rm box}$ denotes the Dirac distribution with support on the walls of the box. 
Thus from (\ref{a6}) we have
\be\label{divj}
\nabla^aT_{ab}=\sigma^a F_{ab} \delta_{\rm box}. 
\ee
The  energy-momentum current associated to the lab-frame ($\xi^a_{\rm lab}=\partial_{ct}^a$) is
\be\label{fff}
j^{\rm lab}_a\equiv -T_{ab} \xi_{\rm lab}^b
\ee
which is not conserved because of the contributions of the boundary degrees of freedom mentioned above. In fact, from Maxwell equations we get
\be\label{notconscurr}
\nabla_a j_{\rm lab}^a=-J^a F_{ab} \xi_{\rm lab}^b =-\sigma^a E^{{\rm lab}}_a \delta_{\rm box}
\ee 
where we used that $\nabla_{a}\xi^{\rm lab}_{b}+\nabla_{b}\xi^{\rm lab}_{a}=0$ because $\xi^{\rm lab}_a$ is a Killing field, recall \eqref{kilab}. Note that the right hand side of the previous equation vanishes inside the box where $J^a=0$. If the box is at rest then $\sigma^a E^{{\rm lab}}_a=0$ on the boundary due to the perfect conductor boundary conditions\footnote{Otherwise the charges would accommodate as they can move due to the electric force and neutralise any  parallel components of the electric field.} and the current is conserved. However, one can have $\sigma^a E^{{\rm lab}}_a\not=0$ in general situations where the box is moving; such possibility is important and plays a role in Section \ref{work}. 
\subsection{Maxwell equations in the accelerated frame of the box}\label{piripipi}

A specially interesting case for the present paper is the one corresponding to a uniformly accelerated box. Thus we analyse the content of Maxwell theory in terms of the electric and magnetic fields defined in an accelerating frame. 
Given the four velocity of a family of observers at rest with respect to the accelerating box $u_{\rm box}^a$--recall \eqref{ubox}--one can write the
electromagnetic field strength as
\be\label{a1}
F_{ab}=-2 E^{\rm box}_{[a} u^{\rm box}_{b]}-\epsilon_{abcd} B_{\rm box}^c u_{\rm box}^{d}
\ee
where $E^{\rm box}_a=F_{ab} u_{\rm box}^b$ and $B^{\rm box}_a=-\frac{1}{2}\epsilon_{abcd} F^{cb} u_{\rm box}^b$ \footnote{Indeed, the previous expression is valid for any timelike vector field $u^a$ of four-velocities representing a field of observers in spacetime.}.
It follows from the skew symmetry of $F_{ab}$ that
\be\label{orto}
B^{\rm box}_au^a_{\rm box}=0=E^{\rm box}_au^a_{\rm box}.
\ee
Now we write Maxwell equations (\ref{m1}) in a way that it would lead to the analog of Gauss and Amp\'ere integral identities for inertial frames but now these are valid in an accelerated frame. This step is rather technical but very important; a general treatment in curved spacetimes can be found in \cite{10.1093/mnras/198.2.339}.

First notice that $u^{\rm box}_a=g_{ab} u^a_{\rm box}=-\x \nabla_a\tau$. This suggests the introduction of a new quantity, $\overline u_a\equiv-\nabla_a\tau$,  which has the following nice properties:
\ba\label{properties}
\nabla_a \overline u^a&=&0\n \\
\nabla_{a} \overline u_b&=& \nabla_{b} \overline u_a\n \\
h^c_a h^d_b\nabla_{c} \overline u_d&=&0,
\ea
where $h_{ab}=g_{ab}+u^{\rm box}_au^{\rm box}_b$ is the spacial metric of the box simultaneity slices, $\tau=$constant slices in Figure \ref{fig1}. The first property says that the $\overline u^a$ congruence is divergence free, 
the second implies that it is surface forming (trivially coming from the fact that $\overline u_a$ is an exact form normal to the $\tau=$constant surfaces), and the last property implies that it is shear free \cite{Wald:1984rg}. 
Equation \eqref{a1} can be written as
\be
F_{ab}=-2 \overline E_{[a} \overline u_{b]}-\epsilon_{abcd} \overline B^c \overline u^{d}
\ee
where $\overline E^a=\x E_{\rm box}^a$ and $\overline B=\x B_{\rm box}^a$. Maxwell equation \eqref{m1} becomes
\ba
-4\pi J_b&=&-\nabla^a(\overline E_{a} \overline u_{b})+\nabla^a(\overline E_{b} \overline u_{a})-\epsilon_{abcd} \nabla^a(\overline B^c \overline u^{d})\\
&=&-(\nabla^a\overline E_{a}) \overline u_{b}-\overline E^{a} \nabla_a \overline u_{b}+ \overline u^{a} \nabla_a \overline E_{b}-\epsilon_{abcd} (\nabla^a \overline B^c )\overline u^{d}\n \\ 
&+&\underbrace{\overline E_{b} \nabla^a \overline u_{a}-\epsilon_{abcd} \overline B^c (\nabla^a \overline u^{d})}_{=0},\n 
\ea
where for the moment we just used the Leibniz rule and wrote at the end the two terms that vanish identically due to the first two identities in \eqref{properties}.
The next step is to separate the previous equation into its part parallel to $u^a_{\rm box}$ (projecting with $-u^a_{\rm box} u^{\rm box}_b$) and its normal or spacial part (which we can obtain by projecting with $h^a_b=\delta^a_b+u_{\rm box}^a u^{\rm box}_b$).  

Before doing the projections we notice that
\ba\label{est}
-4\pi J_b&=&-(\nabla^a\overline E_{a}) \overline u_{b}-\overline E^{a} \nabla_a \overline u_{b}+ \overline u^{a} \nabla_a \overline E_{b}-\overbrace{\epsilon_{abcd} (\nabla^a \overline B^c )\overline u^{d}}^{{\rm orthogonal\ to} \ \overline u^a}\n\\
&=&-(\nabla^a\overline E_{a}) \overline u_{b}-\overline E^{a} \nabla_b \overline u_{a}+ \overline u^{a} \nabla_a \overline E_{b}-\epsilon_{abcd} (\nabla^a \overline B^c )\overline u^{d}\n\\
&=&-(\nabla^a\overline E_{a}) \overline u_{b}+\overline u^{a} \nabla_b \overline E_{a}+ \overline u^{a} \nabla_a \overline E_{b}-{\epsilon_{abcd} (\nabla^a \overline B^c )\overline u^{d}}\ea
where in the second line we used the second equation in \eqref{properties} for the second term, and in the third line we used that $\overline E_a \overline u^a=0$, or \eqref{orto}. Let us now project along $u_{\rm box}^a$ recalling that $u^a_{\rm box} \overline u_a=u^a_{\rm box} (u^{\rm box}_a\x^{-1})=-\x^{-1}$ we get
\ba
-4\pi \x J_b u^b_{\rm box}&=&(\nabla^a\overline E_{a}) +2 u_{\rm box}^{a}u_{\rm box}^{b} \nabla_b \overline E_{a} \n \\
&=& (g^{ab}+u_{\rm box}^a u_{\rm box}^b)\nabla_b \overline E_{a}+ u_{\rm box}^{a}u_{\rm box}^{b} \nabla_b \overline E_{a}\n\\
&=& \underbrace{(g^{ab}+u_{\rm box}^a u_{\rm box}^b)\nabla_b \overline E_{a}}_{\equiv D^a\overline E_{a}}- (u_{\rm box}^{b} \nabla_b u^{\rm box}_{a}) \overline E^{a},
\ea 
where  in the last line we used $\overline E_a \overline u^a=0$ again and we have used the definition of the spacial covariant derivative $D_a$ such that $D_a h_{bc}=0$ \cite{Wald:1984rg}. Substituting the expression \eqref{acele} of the acceleration, and $\overline E_a=\x E_a^{\rm box}$ in the last equation we obtain the familliar Gauss law
\be
-4\pi J_b u^b_{\rm box}=\frac{1}{\x}D^a (\x E_a^{\rm box})-\frac{D_a \x}{\x}  E^a_{\rm box} 
\ee
simplifying
\be\label{gausslaw}
\boxed{-4\pi  J_a u_{\rm box}^a =D^a (E^{\rm box}_a),}
\ee
which has the form of the usual Gauss law in an inertial frame. Indeed it is easy to show that the Gauss law holds in its usual form in arbitrary frames 
(see Problem 2 in Chapter 4 of \cite{Wald:1984rg}). In the present case the technical complications of the previous lines are justified not by the objective of obtaining the Gauss law but rather the aim of getting the analog of Amp\'ere's law which will follow from the spacelike part of the previous equations.

Therefore, we need to project \eqref{est} using $h^a_b=\delta^b_c+u_{\rm box}^b u^{\rm box}_c$. 
But before we notice that the first term projects to zero while the last term projects to itself. 
Let us analyse the remaining terms before projecting. There is 
\be
\overline u^{a} \nabla_b \overline E_{a}=-\overline E^{a} \nabla_b \overline u_{a},
\ee
which in its form on the right clearly projects to zero according to the third equation in \eqref{properties} and the fact that $\overline E^a$ is purely spacelike.
Now let us analyse the remaining term
\be
 \overline u^{a} \nabla_a \overline E_{b}= \x^{-1} (\underbrace{u_{\rm box }^{a} \nabla_a \overline E_{b}+ \overline E_{a} \nabla_b u_{\rm box}^{a}}_{\sL_{u_{\rm box}} \overline E_b}-\overline E_{a} \nabla_b u_{\rm box}^{a}), 
\ee
where we have added and subtracted the same term on the right just to recover the expression of the Lie derivative $\sL_{u_{\rm box}} \overline E_b$ which is the natural derivative along the world-lines of the box observers. Notice that the term we added and subtracted projects to zero (purely time-like) due to the third equation in \eqref{properties}.  The Lie derivative in the previous equation corresponds to a natural proper time ${\rm T}\equiv \x \tau$ derivative of the electric $\overline E_a$. Its space projection is the proper time Fermi transport derivative \cite{10.1093/mnras/198.2.339},  we denote this
\be
D_{\rm T} \overline E_a\equiv h_a^b (\sL_{u_{\rm box}} \overline E_b)=h_a^b ( \overline u^{c} \nabla_c \overline E_{b}), 
\ee
where the previous equivalence of derivatives is valid in our simple case due to \eqref{properties}. For the general relationship among these see \cite{10.1093/mnras/198.2.339}.
Thus, finally putting all this together and projecting into the space part of \eqref{est} we get
\be\label{ampere}
\boxed{-4\pi \x J^{\rm space-part}_b=D_{\rm T} (\x E^{\rm box})_{b}-(D \times \x B^{\rm box})_b},
\ee
where (as in \ref{gausslaw}) $D_a$ is the 3d covariant derivative compatible with the space metric $h_{ab}$.
Finally, the homogeneous Maxwell equations \eqref{m2} can be written as
\be\label{kiki}
\nabla^a F^{\star}_{ab}=0
\ee
where \be F^{\star}_{ab}=\frac{1}{2} \epsilon_{abcd} F^{cd}=2 B^{\rm box}_{[a} u^{\rm box}_{b]}+\epsilon_{abcd} E_{\rm box}^c u_{\rm box}^{d}. \ee
The previous is the analog of $F_{ab}$ as given in \eqref{a1} where $B_a\to -E_a$. As $B_au^a_{\rm box}=0$ as well and this was the only requirement entering the derivation of \eqref{ampere} and \eqref{gausslaw} in addition to the properties of $\overline u_a$ \eqref{properties}, it follows  from \eqref{kiki} that
\be\label{m3}
\boxed{D^a (B^{\rm box}_a)=0,}
\ee
and 
\be\label{m4}
\boxed{-D_{\rm T}(\x B^{\rm box})_{b}+(D \times \x E^{\rm box})_b=0.}
\ee
Equations \eqref{ampere}, \eqref{gausslaw}, \eqref{m3}, and \eqref{m4} are Maxwell's equations for the electric and magnetic field on the accelerated (instantaneous rest) frame of the confining box. 

\subsection{Consequences}\label{conse}

In the previous section we have recast the Maxwell equations in terms of the electric and magnetic fields as measured in the rest frame of the accelerating box. It was a bit technical but the consequences for the electromagnetic field near the perfectly conducting walls of the box are quite simple and analogous to those that one is familiar with for a box at rest in an inertial frame. In this short section we analyse and state them. { We will now see that, as in the case of an inertial box, the Maxwell equations (plus the standard physical assumption that the magnetic field inside the conductor is initially zero) applied to the accelerating box imply that \be\label{fifa} F_{ab}({\rm inside\ conductor})=0.\ee  More precisely, the idealization of perfectly conducting walls requires first the electric field to vanish inside the conductor, $E^{\rm box}_{a}({\rm inside\ conductor})=0$. In addition, right inside the box and at the walls any parallel component of $E^{\rm box}_{a}$ to the walls must vanish: if not there would be a force rearranging surface charges to make this component vanish \footnote{Here we are assuming idealized charge carriers without mass. Real massive charges would produce a parallel $E^{\rm box}_{a}$ component to equilibrate for the gravitational pull as it is intuitive from the perspective offered by the right panel in Figure \ref{photon2}.}}. 
Therefore $E^{\rm box}_{a}|_{\rm box}\propto N_a$ where $N_a$ is the normal to the walls. Now, equation \eqref{m4} implies that the magnetic field $B_a^{\rm box}$ must be time independent inside the conductor. Assuming that the magnetic field was zero initially then we have that $B^{\rm box}_{a}({\rm inside\ conductor})=0$ for all times. Equation \eqref{fifa} now follows from \eqref{a1}. 

Another important consequence of the vanishing of the magnetic field inside the box is that from \eqref{m3} one can prove that the normal component of $B^{\rm box}_{a}$ at the walls (on the inside of the box) must vanish.  An important consequence of this follows from a two lines calculation that uses \eqref{a1}, the definition $j_a^{\rm box}\equiv -T_{ab} u^b_{\rm box}$, and \eqref{jijo}, and leads to the important equation 
\be\label{ortoto}
\left. j^{\rm box}_aN^a\right|_{\rm walls}=\frac{1}{4\pi}\left. \vec B_{\rm box}\cdot (\vec E_{\rm box} \times \vec N)\right|_{\rm walls}=0,
\ee
where we have used once again that $E^{\rm box}_{a}|_{\rm box}\propto N_a$.

In the two sections that follow we will use \eqref{fifa} and Maxwell's equations to express $F_{ab}$ at the walls of the box explicitly in terms of the surface charge current. This will then allow us to write explicitly the energy-momentum tensor \eqref{jijo} at the walls which is important in the analysis of Appendix  \ref{work}. We do this first using three-dimensional methods that involve Gausses law and Amp\`ere's law and later in a more direct covariant fashion.

\subsection{Electromagnetic field at the boundary: canonical derivation}

Let us consider the case of a box made of perfectly conducting walls. Then the presence of surface charges is characterized by the four-current 
\be\label{bibibox}
J^a=\sigma^a \delta_{\rm box}
\ee 
As charges can move freely on the walls, only the 
normal component of the electric field is non vanishing at the wall.  One can use this fact and the Gauss law (the integral form of \eqref{gausslaw} using a suitably chosen region) at the wall and obtain
\be
\left. E^{\rm box}_a\right|_{\rm walls}=-4\pi \left. (\sigma_b u^b_{\rm box}) \right|_{\rm walls} N_a,
\ee
where $N^a$ is the unit normal to the wall.   
Similarly, using Amp\`ere's law  (Stokes theorem and \eqref{ampere}), and the fact that the magnetic field has only parallel components to the walls, one obtains 
\be\label{rero}
\left. B^{\rm box}_a\right|_{\rm walls}=-\left.4\pi \epsilon_{abcd} \sigma^bN^c u^{d}_{\rm box}  \right|_{\rm walls}.
\ee
In order to prove the previous statement one chooses an infinitesimal 2-surface transversal to the walls and such that its normal is aligned with the surface current.
This choice and equation  \eqref{ampere}  yields immediately \eqref{rero}; the time derivative term in  \eqref{ampere} does not contribute because it is orthogonal to the surface's normal.
 
With this the field-strength \eqref{a1} on the walls of the box is given by
\ba\label{efe}
\boxed{F_{ab}
= 8 \pi N_{[a}\sigma_{b]}.}
\ea

\subsection{The energy momentum tensor at the walls}

With the previous result we can now write the energy momentum tensor at the walls of the box using its definition
\eqref{jijo} 
\ba\label{teto}
\left.T_{ab}\right|_{\rm box}
&=&{4 \pi}\left(\sigma_a\sigma_b+(\sigma\cdot \sigma) N_aN_b-\frac{g_{ab}}{2} (\sigma\cdot \sigma) \right),
\ea
where we have used the boundary condition $N^a\sigma_a=0$. It follows that

\be\label{alpha}
\left.j^{\rm lab}_aN^a\right|_{\rm box}=-2\pi (\sigma\cdot \sigma)  N_b\xi_{\rm lab}^b,
\ee
and from \eqref{notconscurr} and \eqref{efe}
\be\label{beta}
\nabla^aj^{\rm lab}_a=\left.- \sigma^a F_{ab} \xi_{\rm lab}^{b}\right|_{\rm box} \delta_{\rm box}=4\pi (\sigma\cdot \sigma)  N_b\xi_{\rm lab}^b \delta_{\rm box}.
\ee
Finally, notice that the pressure at the walls is given by
\be
P_N\equiv T_{ab} N^aN^b=2\pi (\sigma\cdot \sigma) .
\ee


\subsection{Electromagnetic field at the boundary: covariant derivation}

\begin{figure}
\centerline{\hspace{0.5cm} \(
\begin{array}{c}
\includegraphics[width=8cm]{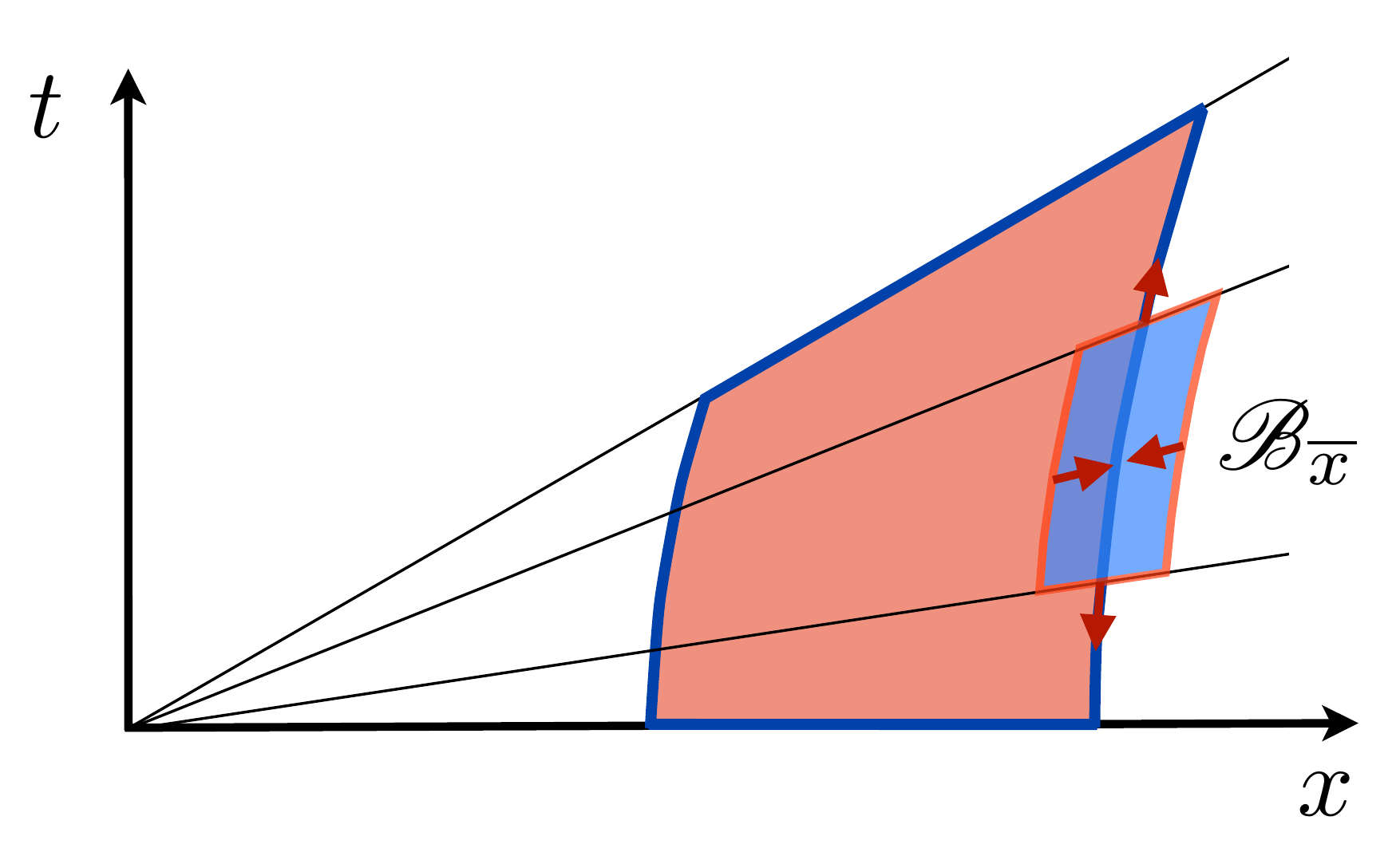} 
\end{array}\) } \caption{Four-dimensional region $\sB_\x$ at the boundary of the world-tube of the box 
where the application of Gauss theorem leads directly to \eqref{efe}. }
\label{ampere-fig}
\end{figure}

Equation \eqref{efe} was derived from the Gauss and Amp\`ere's  laws on an accelerating box in a way that follows the standard text books type of considerations in inertial frames that break covariance by invoking the electric and magnetic fields. However, the simplicity of \eqref{efe} calls for a more direct and covariant derivation. As an exercise, here we show that such more direct path is actually available. 

The key input is the requirement that charges on the wall move freely and so cancel any parallel component of the electric field $E^{\rm box }_a=F_{ab} u^{b}_{\rm box}$
on the rest frame of the box, and similarly that  the normal magnetic field component vanishes at the wall (as shown in Appendix \ref{conse}). 
Consider the four vectors $X^a\equiv \nabla^a \x$, $Y^a\equiv \nabla^a y$, $Z^a\equiv \nabla^a z$, and $T^a\equiv \nabla^a \tau$ which by definition are such that
\be
\nabla^{[a}X^{b]}=\nabla^{[a}Y^{b]}=\nabla^{[a}Z^{b]}=\nabla^{[a}T^{b]}=0.
\ee
The previous set of equations together with the definition of the currents $p^Y_a= F_{ab} Y^b$, $p^Z_a= F_{ab} Z^b$, and $p^T_a= F_{ab} T^b$ imply, from \eqref{m1}, that
\be
\nabla^ap^Y_a=-4\pi J_aY^a,\ \ \ \ \nabla^ap^Z_a=-4\pi J_aZ^a, \ \ \ \ \nabla^ap^T_a=-4\pi J_aT^a.
\ee
Applying the Gauss theorem to $p^Y_a$ in the region $\sB_\x$ shown in figure \ref{ampere-fig} we get
\ba
\int_{\sB_\x} \nabla^ap^Y_a&=&\int_{\partial \sB_\x} p^Y_a n^a \n \\
-4\pi\int_{\sB_\x}  J_a Y^a&=&\int_{\partial \sB_\x} F_{ab} n^aY^b,
\ea
where $n^a$ is the normal to the boundary $\partial \sB_\x$ with the orientation shown in Figure \ref{ampere-fig}.
Notice that as the  normal to the bottom and top (spacelike) portions of the boundaries are proportional to $u^a_{\rm box}$ one has there
that $F_{ab}n^aY^b\propto E^{\rm box}_a Y^a=0$ as only the normal component along $X^a$ of the rest-frame electric field $E^{\rm box}_a$ is non vanishing due to the presence of perfectly conducting walls. In addition $F_{ab}=0$ on the right piece of the timelike component of $\partial \sB_\x$. Therefore, only the integral on the left timelike piece contributes to the right hand side of the previous equation. From this and equation \eqref{bibibox}, together with the assumption that the region $\sB_\x$ is infinitesimally thin around the wall, we get
\be
-4\pi\int d\tau dy dz \x (\sigma_a Y^a)=\int  d\tau dy dz \x  (F_{ab} X^aY^b),
\ee
where we used that the volume density is $\x d\tau dy dz$ and that the normal (oriented for the Gauss theorem is) $n^a=X^a$; however, the conventional inner pointing normal of the box $N^a$.
Assuming that the region is infinitesimal in all directions the previous identity implies
$F_{ab} X^aY^b=-4\pi\sigma_a Y^a$ which in terms of $N^a$ reads
\be\label{eins}
F_{ab} N^aY^b=4\pi\sigma_a Y^a.
\ee
The same logic applied to the current $p^Z_a$ implies
\be\label{zwei}
F_{ab} N^aZ^b=4\pi\sigma_a Z^a.
\ee
A moment of reflection shows that the argument is also true for the current $p_a^T$. Now the top and bottom contributions vanish because the normal there $n^a\propto T^a$ and thus $F_{ab}n^aT^a=0$ because of the skew symmetry of $F_{ab}$. Therefore, we also have
\be\label{drei}
F_{ab} N^aT^b=4\pi\sigma_a T^a.
\ee
The most general $F_{ab}$ would be of the form
\be
F_{ab}=f_{TN} T_{[a}N_{b]}+f_{TY} T_{[a}Y_{b]}+f_{TZ} T_{[a}Z_{b]}+f_{YN} Y_{[a}N_{b]}+f_{ZN} Z_{[a}N_{b]}+f_{YZ} Y_{[a}Z_{b]}.
\ee 
The fact that the electric field $E^{\rm box}_a$ is proportional to $N^a$ due to the presence of the wall implies that $f_{TY}=f_{TZ}=0$.
As the normal component of the magnetic field must vanish due to the present of the conducting wall we also have $f_{YZ}=0$. Thus
\be
F_{ab}=f_{TN} T_{[a}N_{b]}+f_{YN} Y_{[a}N_{b]}+f_{ZN} Z_{[a}N_{b]}.
\ee 
Equations \eqref{eins}, \eqref{zwei}, and \eqref{drei} fix the last three components. The solution is
\be
\boxed{F_{ab}=8\pi N_{[a} \sigma_{b]}},
\ee
which is the same as \eqref{efe}.
One can easily check that the same solution follows from the same argument applied to regions $\sB_{y}$ and $\sB_z$ adapted to the world sheets of the other walls of the box.

\section{Work done by the walls (Maxwell case)}\label{work}

We have seen that the energy content of the box, as measured in the lab frame, depends on $\tau$.  This is due to the action of an external agent that is accelerating the box of radiation. The change in the energy $E(\tau)$  is thus related to the work  done by the external agent on the box. This is associated with the failure for the current $j_a$ to be conserved: in our idealization of the accelerating box, the external agent acts upon the electromagnetic field  via the boundary charges that impose the box boundary conditions and source the divergence of $j_a$ (see \eqref{notconscurr}). The dynamical contribution of these charges is feeding energy into the system. In fact, from the Gauss law, now applied to $j_a$ in the region of interest (Figure \ref{fig1}) we get
\ba\label{tres}
E(\tau)-E(0)=\Delta W &\equiv& \int_{R} \nabla^a j^{\rm lab}_a -\int_{\partial R-\Sigma_1-\Sigma_2} j^{\rm lab}_a N^a \n \\
&=&2\pi\int_{\partial R-\Sigma_1-\Sigma_2}(\sigma\cdot\sigma) N_b\xi_{\rm lab}^b \label{emi}
\ea
where we have used (\ref{alpha}) and \eqref{beta} and the Gauss theorem where the bulk integral involves the integration of the $\delta_{\rm box}$ distribution whose support is at the boundary of $R$. Now from \eqref{tritri} we observe that  $N_b\xi^{b}_{\rm lab}=0$ on any parts of the boundary where $\partial_{\bar x}$ is tangent to the boundary. At the bottom and at the top we have $N_b\xi^{b}_{\rm lab}=\mp \gamma(v) v$ (where for simplicity we are assuming that the box is a cube with walls defined by $\bar x$, $y$, and $z$ equal constant). Therefore, using this in the last line of \eqref{emi}  we get \be
E(\tau)-E(0)=\left(\int_{\rm top}-\int_{\rm bottom}\right)2\pi\gamma v (\sigma\cdot\sigma).
\ee
On the other hand, the pressure on the top/bottom is given by
\ba
P_{\bar x}\equiv T_{ab}\partial_{\bar x}^a\partial_{\bar x}^b=T_{\bar x\bar x}=2\pi (\sigma\cdot\sigma).
\ea
So we find the following expression for the work
\ba
E(\tau)-E(0)&=&\left(\int_{\rm top}-\int_{\rm bottom}\right)P_{\bar x}\bar x\gamma v d\tau dydz\\
&=&\int \sinh\tau d\tau \left.\left( \bar x \int P_{\bar x}dydz\right)\right|_{\rm bottom}^{\rm top} .
\ea
This expression tells us that the origin of the inertia of the box (its resistance to acceleration encoded in the mass \eqref{res}) is the difference of the radiation pressure of the electromagnetic field on the walls between the top and the bottom of  the box.
In order to accelerate the box, an external agent must impose an external force to compensate for the radiation pressure of the confined radiation. Its infinitesimal version
is
\be
 \frac{d E}{d \tau}=\gamma v  \left.\left( \bar x \int P_{\bar x}dydz\right)\right|_{\rm bottom}^{\rm top}. 
\ee
Let us define 
\be
F^{\bar x}= \int P_{\bar x}dydz
\ee
In order to better interpret the previous result let us introduce the proper time measured at the center of the box ${\rm T_c} =\bar x_{\rm c} \tau$,  and recall that we denote by $L$ the legth of the box. With this the previous equation becomes
\be
 \frac{{d} E}{d {\rm T}_c}=\gamma v  \left( \left(1+ \frac{ L}{2\bar x_{\rm c}}\right)F^{\bar x}_{\rm top}-\left(1- \frac{L}{2\bar x_{\rm c}}\right) F^{\bar x}_{\rm bottom}\right),
\ee
and using that the magnitude of the acceleration of the center of the box is (according to \eqref{ace}) $|a|=c^2/\bar x_c$ we arrive at
\be
 \frac{{d} E}{d {\rm T}_c}=\gamma v  \left( F^{\bar x}_{\rm top}-F^{\bar x}_{\rm bottom}\right)+ \frac{|a| L}{2 c^2}\gamma v \left( F^{\bar x}_{\rm top}+F^{\bar x}_{\rm bottom}\right). 
\ee
Let us define $F_{\rm net}\equiv \left( F^{\bar x}_{\rm top}-F^{\bar x}_{\rm bottom}\right)$ for what follows. If both the length and acceleration of the box are small in the sense that by the time light travels the distance $L$ the velocity increase of the box due to the acceleration is much smaller than $c$ then 
\be
\frac{|a| L}{2 c^2} \ll 1.
\ee
For such small-size/small-acceleration boxes (those that model well a composite particle) we recover the usual relativistic
law
\be
 \frac{{\rm d} E}{\rm d T_c}=\gamma v F_{\rm net}. 
\ee
or in covariant notation $u_{\rm c}^a\nabla_a{E}= F^a_{\rm net} \xi^{\rm lab}_a$. The previous equation implies the familiar second Newton equation in the instantaneous rest frame 
\be
m a=F_{\rm net}
\ee
where $m$ is given by $E(0)/c^2$ as given in \eqref{energia}. This shows that the physical origin of mass can be traced 
to the inertia produced by the difference of radiation pressure between the top and the bottom of the box (in analogy with the heuristic simplistic picture given in terms of the bouncing photon in Section \ref{photonh}).

\section{Work done by the walls (scalar field case)}\label{workscalar}

In this section we repeat the derivation of the previous one but in the case of the massless scalar field.
In the lab frame  $\xi_{\rm lab}^a=\partial_t^a$  the associated energy-momentum current is
\be\label{fff}
j^{\rm lab}_a\equiv -T_{ab} \xi_{\rm lab}^b
\ee
which, as in the electromagnetic case,  is not  conserved due to the contributions of the boundary degrees of freedom. One has
\be\label{notconscurr-s}
\nabla_a j_{\rm lab}^a=-\sigma \,  u^b \nabla_b\phi \delta_{\rm box} 
\ee 
where we used \eqref{kilab}. 
The energy momentum tensor \eqref{emt} at the walls is therefore
\be
\left.T_{ab}\right|_{\rm box}= \sigma^2 \left(N_aN_b-\frac12 g_{ab}\right) 
\ee
Now from \eqref{divtabp} we have
\be\label{divtaby}
\nabla^aT_{ab}=\sigma^2 N_b \delta_{\rm box},
\ee
from \eqref{fff}
\be\label{alphay}
\left.j_aN^a\right|_{\rm box}=-\frac12\sigma^2 N_bu^b,
\ee
and from \eqref{notconscurr-s}
\be\label{betay}
\nabla^aj_a=- \sigma^2 N_bu^b \delta_{\rm box}.
\ee
Finally, notice that the pressure at the walls is given by
\be\label{pepe}
P_N\equiv T_{ab} N^aN^b=\frac{1}{2} \sigma^2.
\ee
As in the electromagnetic case, equations \eqref{divtaby}, \eqref{alphay}, \eqref{betay}, and \eqref{pepe} and the same line of argument of Section \ref{work} lead to
\be
 \frac{d E}{d \tau}=\gamma v  \left.\left( \bar x \int P_{\bar x}dydz\right)\right|_{\rm bottom}^{\rm top}. 
\ee
The same conclusions as for the Maxwell case follow from here.

\end{appendix}
\providecommand{\href}[2]{#2}\begingroup\raggedright

\endgroup

\end{document}